\documentclass[aps,prl,reprint,superscriptaddress,showpacs,amssymb,floatfix]{revtex4-1}
\usepackage[final]{graphicx}

\begin{document}

\title{The 3D Lima\c{c}on: Properties and Applications}

\author{Jakob Kreismann}
\email{jakob.kreismann@tu-ilmenau.de}
\affiliation{Institute for Physics, Group for Theoretical Physics II / Computational Physics, Technische Universit\"{a}t Ilmenau,  Weimarer Stra\ss{}e 25, 98693 Ilmenau, Germany}

\author{Stefan Sinzinger} 
\affiliation{Department of Mechanical Engineering, Optical Engineering Group, Technische Universit\"{a}t Ilmenau, Helmholtzring 1, 98693 Ilmenau, Germany}

\author{Martina Hentschel}
\affiliation{Institute for Physics, Group for Theoretical Physics II / Computational Physics, Technische Universit\"{a}t Ilmenau,  Weimarer Stra\ss{}e 25, 98693 Ilmenau, Germany}

\date{\today}

\begin{abstract}
We perform electromagnetic wave simulations of fully three-dimensional optical Lima\c{c}on-microcavities, one basis for their future applications in microlasers and photonic devices.
The analysis of the three-dimensional modes and far-fields reveals an 
increase of the quality factors as compared to the two-dimensional case. The 
structure of the far-field in the third dimension shows pronounced maxima in the emission directionality inclined to the resonator plane which may be exploited for coupling the resonator modes to the environment. This triggers ideas for technical applications, like the suggested 
sensor that can detect small changes in the environment based on changes in the emission profile. 
\end{abstract}


\maketitle


The confinement and manipulation of light using microcavites has attracted a lot of interest in basic and applied physics research over the past decades~\cite{Vahala2003,Chang1996}, e.g. research on microlasers~\cite{Levi1992}, filters for communication technology~\cite{McAulay2009} or single molecule sensing~\cite{Armani2007}. Furthermore, the research on micro-combs~\cite{Cao2014,Loures2015,Karpov2016} and optomechanics~\cite{Chen2013,Aspelmeyer2014} benefits from the  progresses made in the field of optical microresonators.  Established examples of optical microcavites are microdisks~\cite{Levi1992,APL90_2007}, microspheres~\cite{Collot1993,Gorodetsky2000} and microtoroids~\cite{Ilchenko2001,Armani2003} which confine light in whispering gallery modes with high quality factors $Q$. The first microdisk-based microlasers had the drawback of isotropic light emission because of rotational symmetry. In order to observe a directional laser emission, deformed microcavities were investigated~\cite{BowTie,Schwefel2004,PRA73_2006}. A promising shape to combine 
directional emission and high quality factors is the Lima\c{c}on-shape~\cite{PRL100_2008}. Here, ray and wave calculations based on a two-dimensional model system agree very well with the experimentally observed far-field characteristics~\cite{PRALimacon1,Yi2009,APL94_2009,PhysRevARayLimacon}.
 
\begin{figure}
\centering
\includegraphics[width=8.5cm]{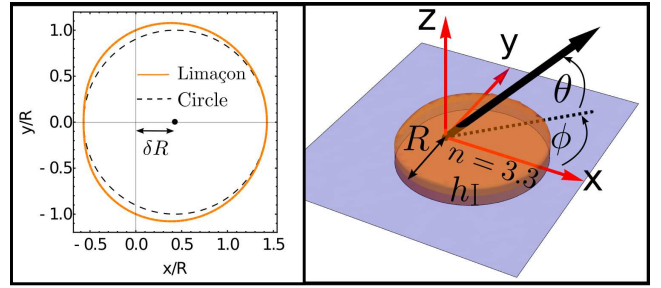}
\caption{Left inset shows the Lima\c{c}on-shape from Eq.~(\ref{equ:2DLimacon}) and for comparison a circle with radius $R$ centred at $x=\delta R$. Note that the origin $(0,0)$ of the polar coordinate system for the Lima\c{c}on is quite off its centre. The right figure displays the 3D Lima\c{c}on cavity with mean radius $R$ and height $h$. $\phi$ is the azimuthal and $\theta$ the inclination angle; $(\phi,\theta)$ determine the far-field direction. The blue plane indicates the resonator plane ($x$-$y$-plane, $z=0$ or $\theta=0$).
}
\label{fig:3DLimacon}
\end{figure}

In reality, however, microcavities are three-dimensional (3D) objects with finite heights. 
This third dimension will be especially important when the cavity sizes are  further reduced and both cavity height $h$ and radius $R$ become comparable to the 
wavelength
\cite{SciRep2014,PRA84_2011}. 
Here, we systematically study 3D microcavities of Lima\c{c}on-shape, see left
inset of FIG.~\ref{fig:3DLimacon}. Its cross section in the $x$-$y$-plane is given in polar coordinates $(r, \phi)$, cf. FIG.~\ref{fig:3DLimacon}, by 
\begin{equation}
r(\phi)=R(1+\delta \cos(\phi) ) \:,
\label{equ:2DLimacon}
\end{equation}

\noindent with mean Radius $R$ and deformation parameter $\delta$. We set 
$\delta=0.43$, a value known~\cite{PRL100_2008} to yield a highly directional far-field emission for two-dimensional (2D) cavities with refractive index $n=3.3$ embedded in vacuum ($n_0=1$) as used here. 
We first discuss modes and far-fields of 3D Lima\c{c}on cavities of varying height to radius ratio $h/R$, followed by outlining a sensor application based on
the 3D character of the far field and its extreme sensitivity to tiny changes in the refractive index. 

Using MEEP~\cite{MEEP}, a free finite-difference time-domain (FDTD) software package, 3D electromagnetic wave simulations have been performed to calculate the normalized frequencies $\Omega=kR=\text{Re}(\omega)R/c$, with $\omega$ being a complex frequency and $c$ the speed of light, the quality factors $Q=-0.5\text{Re}(\omega)/\text{Im}(\omega)$, the distributions of the electric field component $E_{z}(x,y,z)$ (modes) and the far-field intensity $I(\phi,\theta)$. As our focus is on wavelength-scale cavities, $kR$ ranges from $1.9$ up to $10.7$, with $k = 2 \pi / \lambda$ being the wave number and $\lambda$ the wavelength in vacuum. We use a $E_{z}$-point-dipole source to excite the modes and focus on the study of TM-polarized modes.

\begin{figure}
\centering
\includegraphics[width=8.5cm]{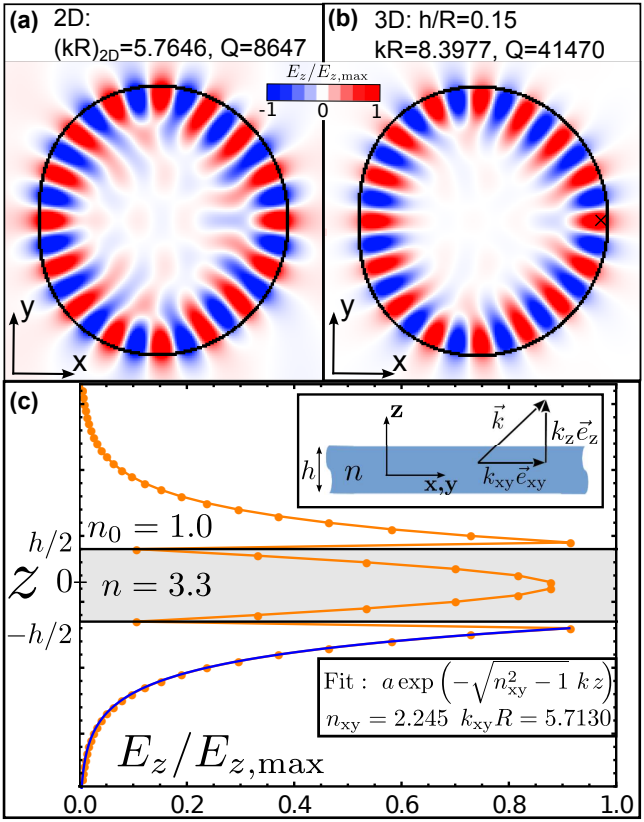}
\caption{3D mode structure. Comparison of (a) a 2D and (b) a 3D Lima\c{c}on mode taken at $z=0$. Both modes have the same radial order $l=1$ and azimuthal order $m=16$. (c) shows $E_{z}$ parallel to the z-axis piercing through the point marked by the cross in (b). Inset in (c) illustrates a dielectric slab and the decomposition of the wave vector $n\vec{k}$. }
\label{fig:compare2D3D}
\end{figure}

We first discuss the analogies between the structures of modes of the 
2D and 3D Lima\c{c}on cavity, respectively.   
An example of a 2D mode and the $(x,y)$-cross section of a 3D mode are shown in FIG.~\ref{fig:compare2D3D}~(a) and (b). Both modes exhibit the same azimuthal order $m=16$ and a similar field distribution of $E_{z}$, but different normalized frequencies $kR$ and quality factors $Q$. The higher $kR$ and the much larger $Q$ of the 3D mode arise from the additional confinement in the third dimension and from the faster fall-off of the electromagnetic field in the resonator ($x,y$) plane ($1/$distance $r$ instead of $1/\sqrt{r}$), respectively, yielding a general increase of $Q$ in finite height cavities which is of crucial experimental relevance (see supplementary material).

In order to investigate the mode structure perpendicular to the resonator plane ($z$-direction),
we analyse $E_{z}(z)$ in FIG.~\ref{fig:compare2D3D}(c) at one $(x,y)$-position marked by a cross in Fig.~\ref{fig:compare2D3D}(b). The mode confinement between the top and bottom surface 
is clearly visible, as well as the expected exponential decay of $E_{z}$ outside this dielectric slab. 
The finite value $E_{z}(z=\pm h/2)$ reflects the boundary condition -- the dielectric displacement field $\vec{D}$ has to be continuous at the top (t) and bottom (b) surface~\cite{Jackson}, 
$(\vec{D}_{\text{out}}-\vec{D}_{\text{in}})\cdot \vec{N}_{\text{t,b}}=0$,
where $\vec{N}_{\text{t,b}}$ is the normal vector at the top and bottom surface, respectively. 
%

We analyse the exponential decay of $E_z(z)$ using 
the so-called effective refractive index model~\cite{neffModel}, cf.~inset of FIG.~\ref{fig:compare2D3D}(c). The key idea is to decompose the 3D wave vector $n\vec{k}$ into its horizontal (index xy) and its vertical (index z) component
\begin{equation}
n\vec{k}=nk_{xy}\vec{e}_{xy}+nk_{z}\vec{e}_{z} \:,
\label{equ:3Dveck}
\end{equation}

\noindent where $\vec{e}_{xy}$ ($\vec{e}_{z}$) is the unit vector in the $x$-$y$-plane (in $z$-direction). 
We make the following ansatz for the exponential decay of  $E_{z}(z)$:
\begin{equation}
 E_z(z) =  \left\{ \begin{array}{lr} 
		  a_1 \exp(ink_{z}z) + a_2 \exp(-ink_{z} z) & : |z| \leq h/2 \\ 
		  a_3 \exp(-q_{z}|z|)                      & : |z| \ge  h/2 	 
			\end{array} ,
		\right.
\label{equ:Ez_Ansatz}
\end{equation}
\noindent where the vertical component $q_{z}$ outside the cavity is related to the horizontal component $nk_{xy}$ inside and the wave number $k$ by $q_{z}^2=(nk_{xy})^2-k^2=k^2(n_{xy}^2-1)$ 
according to the
boundary conditions. 
The effective refractive index $n_{xy}$ in the $x$-$y$-plane 
follows from
the Pythagorean decomposition, 
using  Eq.~(\ref{equ:3Dveck}),
\begin{equation}
 n^2=: n_{xy}^2 + n_{z}^2 = \left(\frac{nk_{xy}}{k}\right)^2 + \left(\frac{nk_{z}}{k}\right)^2 \:.
\label{equ:neff}
\end{equation}

\noindent The $n_{xy}$ represents the ratio of the speed of light in vacuum to that of a horizontally guided mode, and 
is a measure of the inclination angle $\cos(\chi)=n_{xy}/n$ of the 3D wave vector $n\vec{k}$ w.r.t. the $x$-$y$-plane. It runs from $n_{xy}=1$ (total internal reflection on the top and bottom area) up to $n_{xy}=n$ (light propagation in the $x$-$y$-plane). An analytical form of $n_{xy}=n_{xy}(kh)$ for the dielectric slab was derived, e.g., in~\cite{Lebental2007,neffModel,Jackson}.\\
\noindent It is tempting to compare $(kR)_{\text{2D}}$ of the 2D mode with the horizontal component $k_{xy}R$ of the 3D mode. The exponential fit yields $n_{xy}=2.245$ that corresponds to $k_{xy}R=5.7130$. This result is very close to $(kR)_{\text{2D}}=5.7646$ confirming the similar mode structures, as seen in FIG.~\ref{fig:compare2D3D}.

\begin{figure}
\centering
\includegraphics[width=8.5cm]{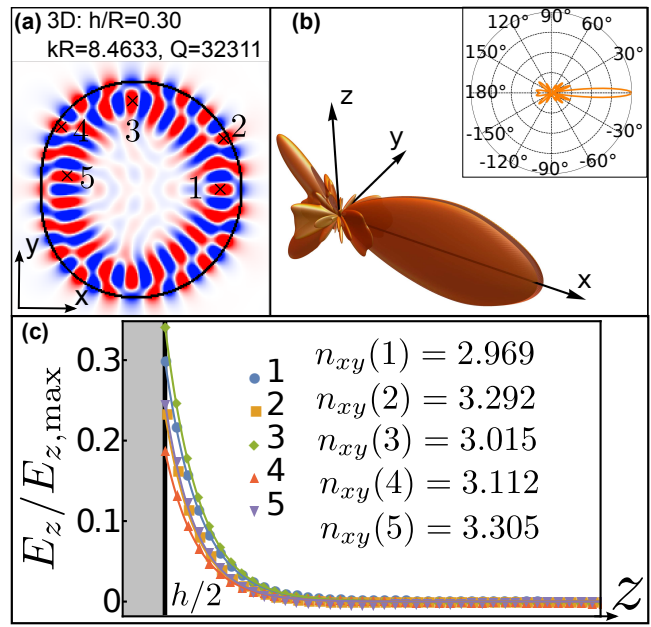}
\caption{Modes and far-fields in 3D of a whispering-gallery type mode. (a) A cross section of a 3D Lima\c{c}on mode at $z=0$. (b) 3D far-field and its cross section in the resonator plane (inset). (c) Electric field component $E_{z}$ above the cavity at positions numbered in (a) and the deduced effective refractive indices $n_{xy}$. The plot markers and solid lines show numerical data and the fit of the exponential decay, respectively. }
\label{fig:mode_ff_1}
\end{figure}
\begin{figure}
\centering
\includegraphics[width=8.5cm]{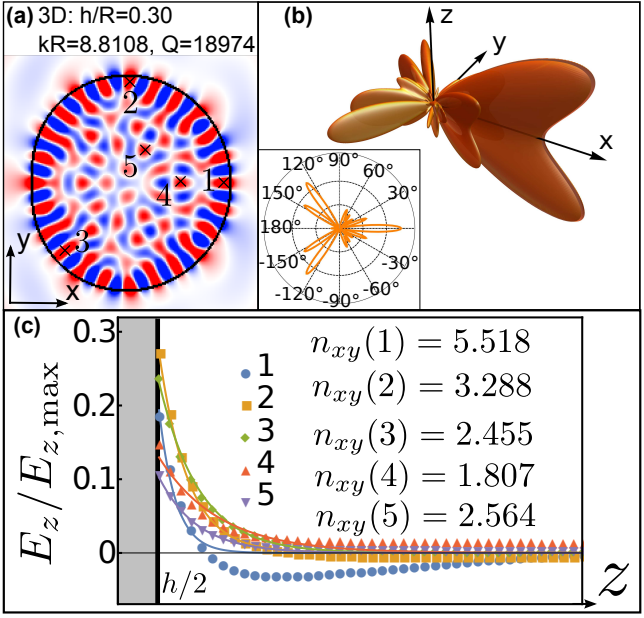}
\caption{As FIG.~\ref{fig:mode_ff_1}, but for a mode characterized by higher field amplitudes in the centre of the cavity. Note the oscillations of $E_z$ in (c), e.g. at position 1. See text for details.}
\label{fig:mode_ff_2}
\end{figure}

We now apply the effective refractive index model in order to investigate the confinement in the third dimension in more detail. To this end, we use numerics to fit $n_{xy}$ from the exponential decay as a function of  the $(x,y)$-position, and make the connection to the 3D far-field. We distinguish between (I) whispering-gallery type modes, as seen in FIG.~\ref{fig:compare2D3D} and (II) modes characterized by higher field amplitudes in the centre of the cavity. We will see below that modes of type II display truly 3D far-field features in contrast to type I, cf. FIG.~\ref{fig:mode_ff_1} and FIG.~\ref{fig:mode_ff_2}.

First (I), we analyse a whispering-gallery type mode, as depicted in FIG.~\ref{fig:mode_ff_1} (a) where the crosses mark the position at which the $n_{xy}$ were fitted from the exponential decay of the electric field outside the cavity, cf.~FIG.~\ref{fig:compare2D3D}. In addition, FIG.~\ref{fig:mode_ff_1} (c) illustrates the electric field  and displays the values of $n_{xy}$ at the marked positions. We clearly observe a similar exponential decay at all positions, denoting a high confinement between the top and bottom area. The resulting $n_{xy}$ vary within a relatively small range from $n_{xy}=2.969$ up to $n_{xy}=3.305$ which indicates that the mode propagates homogeneously and with a large $k_{xy}R$-component since $n_{xy}\approx 3$ corresponds to an inclination angle of the wave vector $n\vec{k}$ w.r.t. the $x$-$y$-plane of about $25^{\circ}$. 

Having analysed the mode propagation, we now make the connection to the 3D far-field of this mode that is depicted in FIG.~\ref{fig:mode_ff_1} (b). It shows a parametric plot of the intensity $I(\phi,\theta)$ in the Fraunhofer region $r\gg2(2R)^2/\lambda$ (far-field) that displays a main lobe centred along the $x$-axis. The intensity of the main lobe is maximal in the plane of the resonator ($x,y$-plane, $\theta=0$) and decays away from this plane. We checked that this behaviour is very similar to that of a plane wave diffracted at a single slit whose width is the cavity height. The inset of FIG.~\ref{fig:mode_ff_1} (b) shows a polar plot of the far-field intensity of this mode in the plane of the resonator that clearly exhibits  directional emission. Thus, we can think of the main components that determine 3D far-field of this mode: 
(i) 
The 2D Lima\c{c}on-shaped cross section of the 3D cavity induces a characteristic, inherent emission profile (directional emission of the Lima\c{c}on).
(ii) 
3D Lima\c{c}on modes experience diffraction at the side area due to its finite height.

Next (II), we focus on modes which exhibit higher field amplitudes in the centre of the cavity, as depicted in FIG.~\ref{fig:mode_ff_2} (a). This mode exhibits a much more complex field distribution inside the cavity as well as outside the cavity, as shown in FIG.~\ref{fig:mode_ff_2} (c). The resulting $n_{xy}$ vary within a wide range from $n_{xy}=1.807$ up to $n_{xy}=3.288$, whereas $n_{xy}=5.518$ is an example of a failed exponential fit because the electric field at position 1 propagates even outside the cavity. This propagation indicates refractive output and as a result of a reduced confinement between top and bottom area, reflected also in a diminished quality factor $Q$. The wide range of observed $n_{xy}$ reflects the complexity of the mode that consists of whispering-gallery-type features at positions 2 and 3, as well as zigzagging (between top and bottom area) features at position 4 since $n_{xy}(4)=1.807$ corresponds to an inclination angle of the wave vector $n\vec{k}$ w.r.t. the $x$-$y$-plane of about $57^{\circ}$.
 
Based on this analysis, we now address the 3D far-field of this mode, shown in FIG.~\ref{fig:mode_ff_2} (b). We observe two main lobes that are inclined w.r.t. the resonator ($x,y$) plane. The inset confirms that directional emission still exists in the ($x,y$)-plane and into the expected direction, but the maximum intensity is emitted along an inclined direction highlighting the 3D character of the far-field. This inclined and directional emission is a truly 3D effect and originates in the output detected around position 1 marked in FIG.~\ref{fig:mode_ff_2} (a). Therefore, for type-II modes we complement the far-field determining mechanisms (i), (ii) discussed above by (iii) refractive escape of electromagnetic waves  through the top and bottom area. 
 
We point out that all far-fields are for free-standing cavities. For a substrate-mounted cavity, additional reflections at the substrate could interfere with the direct light emissions from the cavity and could thus modify the far-fields~\cite{Lafargue2013}.

In the following, we use the truly 3D far-field features described above to design a sensor that can detect particles or gases in the environment based on a change of the emission characteristics, e.g. for a lab-on-a-chip application~\cite{LabChip}. An optical resonator, such as the 3D Lima\c{c}on cavity, can be very sensitive against tiny changes at the boundary. Thus, changes in the environment could affect mode structures~\cite{Armani2003} and consequently the far-fields, too. Since we focus on wavelength-scale cavities, bidirectional emission is possible~\cite{PRA84_2011} unlike in large cavities where universal emission directionality is determined by chaotic ray dynamics. Here, we investigate the extent of the changes in the environment necessary to change the far-field direction significantly.
   
\begin{figure}
\centering
\includegraphics[width=8.5cm]{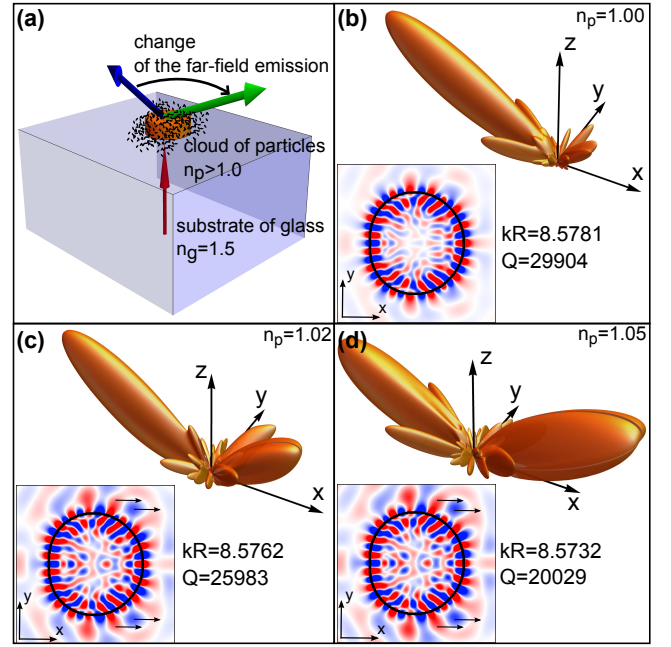}
\caption{3D Lima\c{c}on sensor. (a) sketches the idea of a 3D Lima\c{c}on sensor. (b), (c) and (d) show calculated far-fields and the horizontal mode structures at $z=0$ (insets). Increasing the $n_{\text{p}}$ results in reversed and bidirectional emission in (d) compared to (b).}
\label{fig:3DSensor_FF}
\end{figure}

The idea of a particle-sensor is illustrated in FIG.~\ref{fig:3DSensor_FF}(a). A 3D Lima\c{c}on is placed on a glass substrate that has a typical refractive index $n_{\text{g}}=1.5$. The upper half space is enclosed by a chamber which could contain a gas or cloud of particles with a refractive index $n_{\text{p}}$ higher than the vacuum index of $n_0=1.0$. Initially, we assume the chamber to be vacuum or filled with air (refractive index 1.0).
A laser (red arrow) excites a mode in the 3D Lima\c{c}on, its far-field direction is indicated by the blue arrow. Next, the cavity is exposed to a gas of particles raising the refractive index outside from $n_0=1.0$ to $n_{\text{p}}>1.0$. The change of the refractive index influences the mode and the far-field pattern resulting, e.g., in a reversed far-field emission direction (green arrow) as the sensor measurement signal, cf.~Fig.~\ref{fig:3DSensor_FF}.

Figures~\ref{fig:3DSensor_FF} (b) to (d) show the calculated far-fields and the horizontal mode structures at $z=0$ (insets). Note that the far-fields show exclusively positive inclinations, because we are interested in far-fields in the upper half space. Remarkably, increasing the refractive index $n_{\text{p}}$ by $2~\%$~\cite{nchange} and $5~\%$ (FIG.~\ref{fig:3DSensor_FF} (c) and (d)) completely changes the far-field characteristics. 
It leads to a modified field distribution in the centre of the cavity as well as to higher output indicated by black arrows. Consequently, the quality factor $Q$ decreases successively and an additional far-field lobe appears, pointing in the opposite direction compared to the initial far-field for $n_{\text{p}}=1.0$ in FIG.~\ref{fig:3DSensor_FF} (b).
This allows for a relative measurement of the intensities of the two far-field modes which promises more stable and reliable sensor performance, or alternatively to use one lobe to couple in and out, respectively.

To summarize, we have performed 3D FDTD wave simulations in order to investigate 3D mode structures and 3D far-fields of wavelength-scale 3D optical microresonators that in view of the ongoing miniaturisation of photonic devices will soon become of experimentally relevance.
In particular we show that $Q$ factors can be expected to be higher in 3D systems.
While we find that the mode distribution looks similar inside the cavity for the 2D and 3D cases, the far-fields can differ significantly, implying that 2D far-field simulations can be highly misleading. 

Our results on 3D far-fields embed those obtained for 2D microcavities by explicitly showing how the emission pattern looks perpendicular to the resonator plane. Besides a mere diffractive spreading due to the finite height of the lateral area, we predict a qualitatively new type of far-field where the directions of maximum far-field emission are not in the resonator plane, but inclined to it, forming an angle of almost 60$^o$. It arises for modes with a slightly increased intensity in the cavity centre as compared to whispering-gallery-type modes. This suggests that the far-field properties can be customized by adjusting the mode structure, e.g. via the 3D resonator geometry.



The unique emission characteristics of the 3D microresonators 
can potentially be exploited in sensors and highly integrated optical microsensors~\cite{Kleindienst2015}. We discuss a sensor that can detect tiny changes in the environment via characteristic far-field signal. Furthermore, we plan to utilise integrated optical microsystems to optically address  individual microdisk resonators. The 3D simulations presented here enable the optimization of the coupling efficiencies. Coherently coupled microcavities can be fabricated by carefully aligning arrays of individual resonators which is feasible using lithographic fabrication technologies. This will help to further enhance the sensitivity of the microdisk sensors.

\begin{acknowledgments}
This work was partly supported by Emmy-Noether programme of the German Research Foundation (DFG).
\end{acknowledgments}


%

\end{document}